\documentclass[conference]{IEEEtran}
\IEEEoverridecommandlockouts
\usepackage{cite}
\usepackage{amsmath,amssymb,amsfonts}
\usepackage{algorithmic}
\usepackage{graphicx}
\usepackage{textcomp}
\usepackage{xcolor}
\usepackage{booktabs}
\def\BibTeX{{\rm B\kern-.05em{\sc i\kern-.025em b}\kern-.08em
    T\kern-.1667em\lower.7ex\hbox{E}\kern-.125emX}}
\begin{document}

\title{A Comprehensive Empirical Evaluation of Vector Database Systems for Approximate Nearest Neighbor Search: Performance, Quality, and Resource Trade-offs}

\author{\IEEEauthorblockN{1\textsuperscript{st} 
Ashen Rashmika}
\IEEEauthorblockA{\textit{University of Kelaniya} \\
Sri Lanka\\
ashenrashmike2000@gmail.com}
\and
\IEEEauthorblockN{2\textsuperscript{nd} Tiroshan Madushanka}
\IEEEauthorblockA{\textit{University of Kelaniya} \\
Sri Lanka\\
tiroshanm@kln.ac.lk}
}

\maketitle

\begin{abstract}
Vector databases have emerged as critical infrastructure for modern artificial intelligence applications, particularly retrieval-augmented generation (RAG), semantic search, and recommendation systems. Despite their growing importance, there remains a significant gap in comprehensive, reproducible benchmarks that jointly evaluate retrieval quality, query latency, throughput, and resource utilization. We present a systematic empirical evaluation of seven prominent vector database systems: FAISS, Qdrant, Milvus, Weaviate, Chroma, pgvector, and LanceDB. Our methodology spans six diverse datasets, from classical computer-vision descriptors (SIFT, GIST) to transformer-based text embeddings (MS MARCO, GloVe), encompassing over 4 million vectors at dimensionalities from 96 to 960. We measure 15 metrics spanning retrieval quality (Recall@K, Precision@K, MRR, NDCG@K, Hit Rate@K), query performance (latency percentiles, QPS, cold-start latency), and resource consumption (index build time, memory, storage). On SIFT1M, FAISS achieves the highest single-node throughput (866 QPS) but lacks database operational features; Weaviate provides the best out-of-the-box recall (\textgreater 99\%); Qdrant offers the best latency among full databases (4.55~ms median); and LanceDB trades retrieval quality for substantially faster index construction. We derive system-selection guidelines for practitioners and release our benchmarking framework as open-source software.
\end{abstract}


\section{Introduction}\label{sec:introduction}

Dense vector embeddings, produced by transformers, convolutional networks, and language models, now serve as the universal representation for text, images, audio, and multimodal data \cite{lewis2020retrieval}. This shift from sparse, keyword-based representations to dense semantic embeddings has created unprecedented demand for systems that can store, index, and query billions of high-dimensional vectors with sub-second latency. The rise of large language models (LLMs) and retrieval-augmented generation (RAG) has further accelerated adoption of vector databases, which serve as the external knowledge store grounding LLM responses in semantically retrieved context \cite{lewis2020retrieval}.

Vector databases employ approximate nearest neighbor (ANN) indexing techniques such as Hierarchical Navigable Small World (HNSW) graphs \cite{malkov2018efficient}, Inverted-File indices with product quantization \cite{jegou2011product}, and locality-sensitive hashing \cite{indyk1998approximate} to achieve sub-linear search in high-dimensional spaces. The resulting ecosystem is fragmented: each system makes different architectural trade-offs among retrieval quality, latency, throughput, memory, and operational complexity, and existing benchmarks often evaluate isolated ANN libraries rather than full production database systems.

This paper addresses four research questions: (RQ1) how do systems compare on retrieval quality across diverse datasets; (RQ2) what are their latency and throughput characteristics; (RQ3) what are their index-build and resource requirements; and (RQ4) what trade-offs should guide system selection. Our contributions are: (i) a reproducible, containerized benchmarking framework with pluggable per-system adapters; (ii) an empirical evaluation of seven systems across six datasets, generating over 2,500 individual measurements; (iii) analysis of system-specific strengths, weaknesses, and operational bottlenecks; and (iv) actionable selection guidelines, with the framework released as open-source software.

\section{Background and Related Work}\label{sec:background}

The $k$-nearest-neighbor problem seeks, for a query vector $\mathbf{q}$ and dataset $\mathcal{D}\subset\mathbb{R}^d$, the $k$ vectors minimizing a distance function $\delta$ (typically Euclidean or cosine). Exact search costs $O(nd)$ per query, which is prohibitive at billion-scale $n$ and $d$ in the hundreds, motivating ANN methods that trade exactness for efficiency, measured by recall. Locality-sensitive hashing \cite{indyk1998approximate} offers sub-linear query time via similarity-preserving hash buckets but incurs high memory overhead. Tree-based methods (KD-trees, ball trees) partition space hierarchically but degrade toward linear scan beyond roughly 20 dimensions. Product quantization \cite{jegou2011product} compresses vectors into sub-vector codebooks for memory-efficient approximate distance computation, at some recall cost. Graph-based HNSW \cite{malkov2018efficient} currently dominates production systems, offering logarithmic search complexity, incremental insertion, and tunable recall--latency trade-offs via $M$, efConstruction, and efSearch; DiskANN \cite{subramanya2019diskann} extends this to billion-scale, SSD-backed indexes.

Vector databases extend such ANN algorithms with persistence, concurrent access, metadata filtering, and horizontal scaling. FAISS \cite{johnson2019billion,douze2024faiss} is an optimized library (IVF, HNSW, PQ, optional GPU) without built-in persistence. Qdrant is a Rust-based system emphasizing filtering and horizontal scaling. Milvus is a distributed database separating storage and compute, supporting IVF/HNSW/DiskANN. Weaviate combines HNSW with built-in vectorization and knowledge-graph-style object storage. Chroma targets rapid RAG prototyping with a simple Python API. pgvector extends PostgreSQL for hybrid SQL/vector queries. LanceDB uses a disk-based columnar format with IVF-PQ indexing for fast, memory-efficient construction.

Prior benchmarking efforts are algorithm-centric (ANN-Benchmarks \cite{aumuller2020ann}, Big-ANN-Benchmarks \cite{simhadri2021results}) or emphasize throughput under idealized conditions with limited quality/resource coverage (VectorDBBench \cite{vectordbbench2023}). This work jointly evaluates full production systems across heterogeneous datasets under one controlled environment, reporting quality, latency, and resource metrics together.

\section{Methodology}\label{sec:methodology}

\subsection{Framework and Datasets}
Our containerized benchmarking framework provides pluggable database adapters, dataset loaders with ground-truth computation, metric collectors, and an experiment runner that manages warm-up and result export. For each database--dataset pair the procedure is: (1) initialize the database instance via its adapter; (2) build the index, recording build time, peak memory, and disk usage; (3) run a warm-up phase of 1{,}000 untracked queries to stabilize caches; (4) run the measurement phase, recording per-query latency and retrieved results; (5) compute latency percentiles from the recorded timings; and (6) compute quality metrics by comparing retrieved results against pre-computed ground truth. Each configuration is executed three times, with metrics aggregated as the mean. Table~\ref{tab:datasets} summarizes the six evaluation datasets, spanning computer vision, NLP, and synthetic domains at dimensionalities from 96 to 960 and sizes from 100K to 1M vectors.

\begin{table}[t]
\caption{Benchmark Dataset Characteristics}\label{tab:datasets}
\centering
\begin{tabular}{lccccc}
\toprule
\textbf{Dataset} & \textbf{Vectors} & \textbf{Dim} & \textbf{Dist.} & \textbf{Domain} \\
\midrule
SIFT1M   & 1.0M & 128 & L2  & Vision \\
DEEP1M   & 1.0M & 96  & L2  & Deep learning \\
GIST1M   & 1.0M & 960 & L2  & Vision \\
GloVe    & 0.4M & 100 & Cos & NLP \\
MS MARCO & 0.1M & 768 & Cos & IR \\
Random   & 1.0M & 128 & L2  & Synthetic \\
\bottomrule
\end{tabular}
\end{table}

\subsection{Metrics and Setup}
We report retrieval-quality metrics (Recall@K for $K\in\{1,10,50,100\}$, Precision@10, MRR, NDCG@100, Hit Rate@10), performance metrics (P50/P90/P95/P99 latency, QPS, cold-start latency, index build time), and resource metrics (peak RAM, on-disk index size, bytes/vector, CPU utilization). All experiments ran on a GCP n2-standard-8 instance (6 vCPU, 32~GB RAM, 500~GB SSD, Ubuntu 24.04, Python 3.12.3). Systems used default, out-of-the-box configurations rather than exhaustive tuning, reflecting typical practitioner deployment.

\section{Experimental Results}\label{sec:results}

\subsection{Retrieval Quality}
\begin{figure}[t]
\centering
\includegraphics[width=0.98\linewidth]{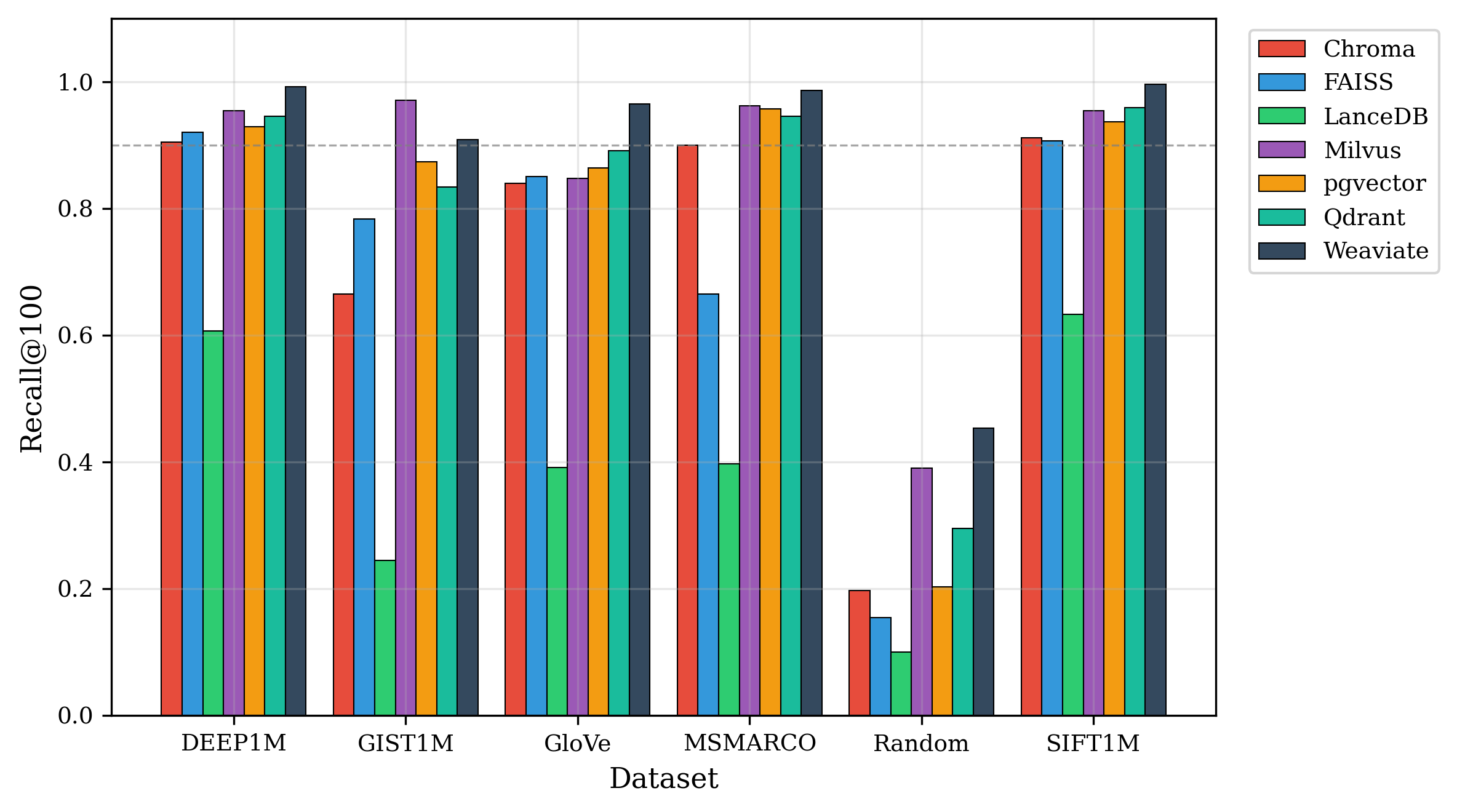}
\caption{Recall@100 across all seven systems and six datasets. The dashed line marks the 90\% recall threshold. Weaviate leads on most datasets; LanceDB trails due to aggressive default quantization.}\label{fig:recall}
\end{figure}

Fig.~\ref{fig:recall} presents Recall@100 across all database--dataset combinations. Weaviate achieves the highest recall on four of six datasets (0.996 on SIFT1M, 0.992 on DEEP1M, 0.965 on GloVe, 0.986 on MS MARCO), reflecting an HNSW configuration tuned for quality. Milvus leads on the highest-dimensional dataset, GIST1M (0.971 at 960D), suggesting its IVF-HNSW hybrid handles high dimensionality well; on that same dataset, LanceDB collapses to 0.245, its steepest drop of any dataset, indicating that its default IVF-PQ codebook is particularly ill-suited to 960-dimensional descriptors. All systems degrade sharply on the unclustered Random dataset (0.10--0.45 recall), confirming that ANN algorithms rely on data structure they cannot exploit in uniform noise; even the strongest system, Weaviate, falls to 0.453 there, a reminder that recall figures measured on clustered, real-world embeddings do not bound worst-case behavior. LanceDB's default IVF-PQ quantization sacrifices 30--70\% recall relative to graph-based systems across every dataset in exchange for memory efficiency and fast index construction, making it the only system in our evaluation whose recall falls below the 0.90 threshold on five of six datasets.

\begin{figure}[t]
\centering
\includegraphics[width=0.85\linewidth]{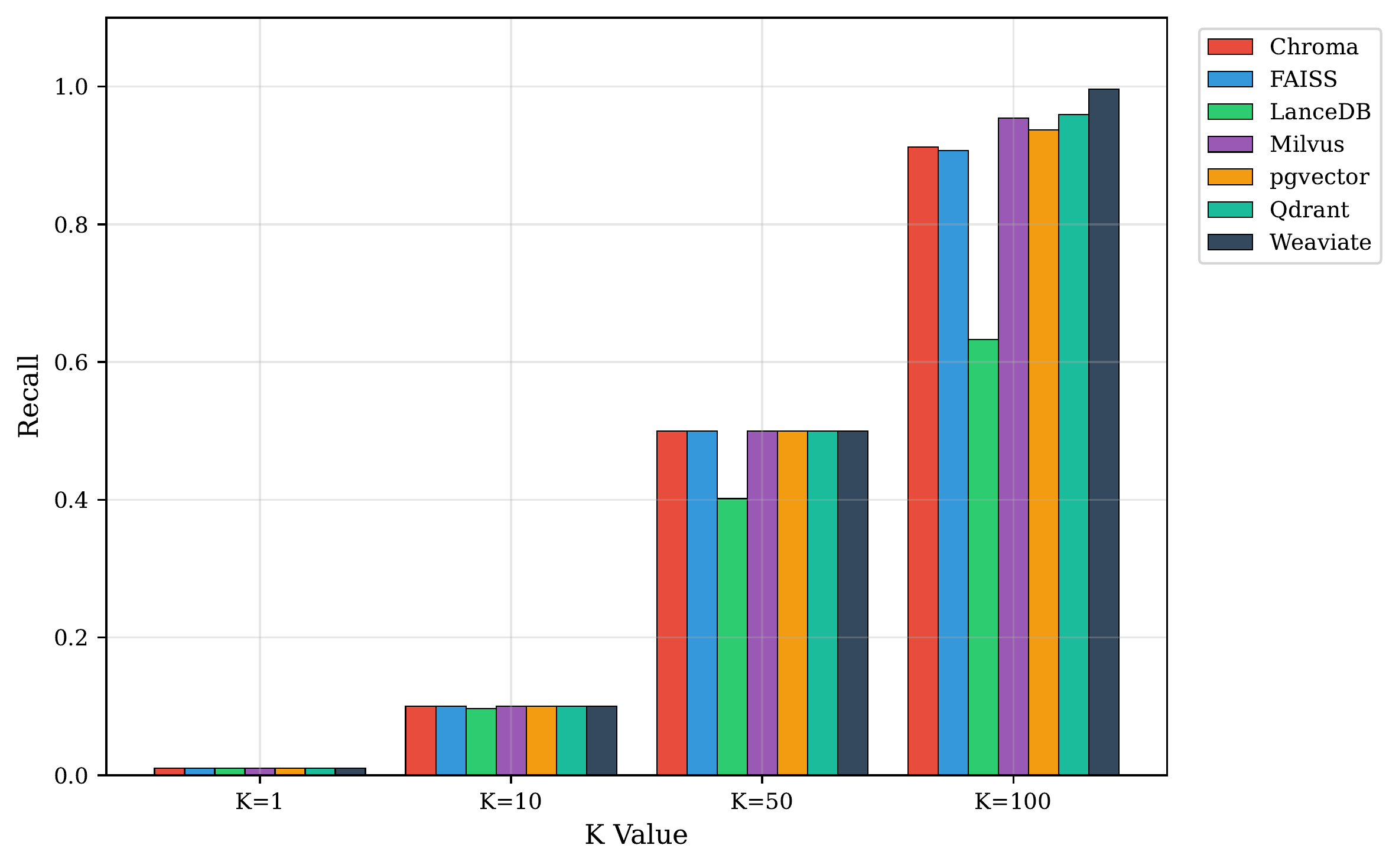}
\caption{Recall at $K\in\{1,10,50,100\}$ on SIFT1M. Systems are nearly indistinguishable at low $K$; separation grows with retrieval depth.}\label{fig:recall_k}
\end{figure}

Fig.~\ref{fig:recall_k} shows how recall varies with the retrieval depth $K$ on SIFT1M. All systems achieve near-perfect Recall@1 and Recall@10 (above 0.98), so shallow retrieval alone cannot distinguish system quality; differentiation emerges only at $K=50$ and $K=100$, where approximate methods must explore a larger portion of the index to maintain quality, exposing the gap between graph-based and quantization-based approaches most clearly. This has a direct practical implication: benchmarks or product evaluations that report only Recall@10 will understate the quality gap that appears once an application (e.g., a RAG pipeline retrieving 50--100 candidate passages for re-ranking) requests deeper result sets.

\subsection{Ranking Quality}
\begin{figure}[t]
\centering
\includegraphics[width=0.98\linewidth]{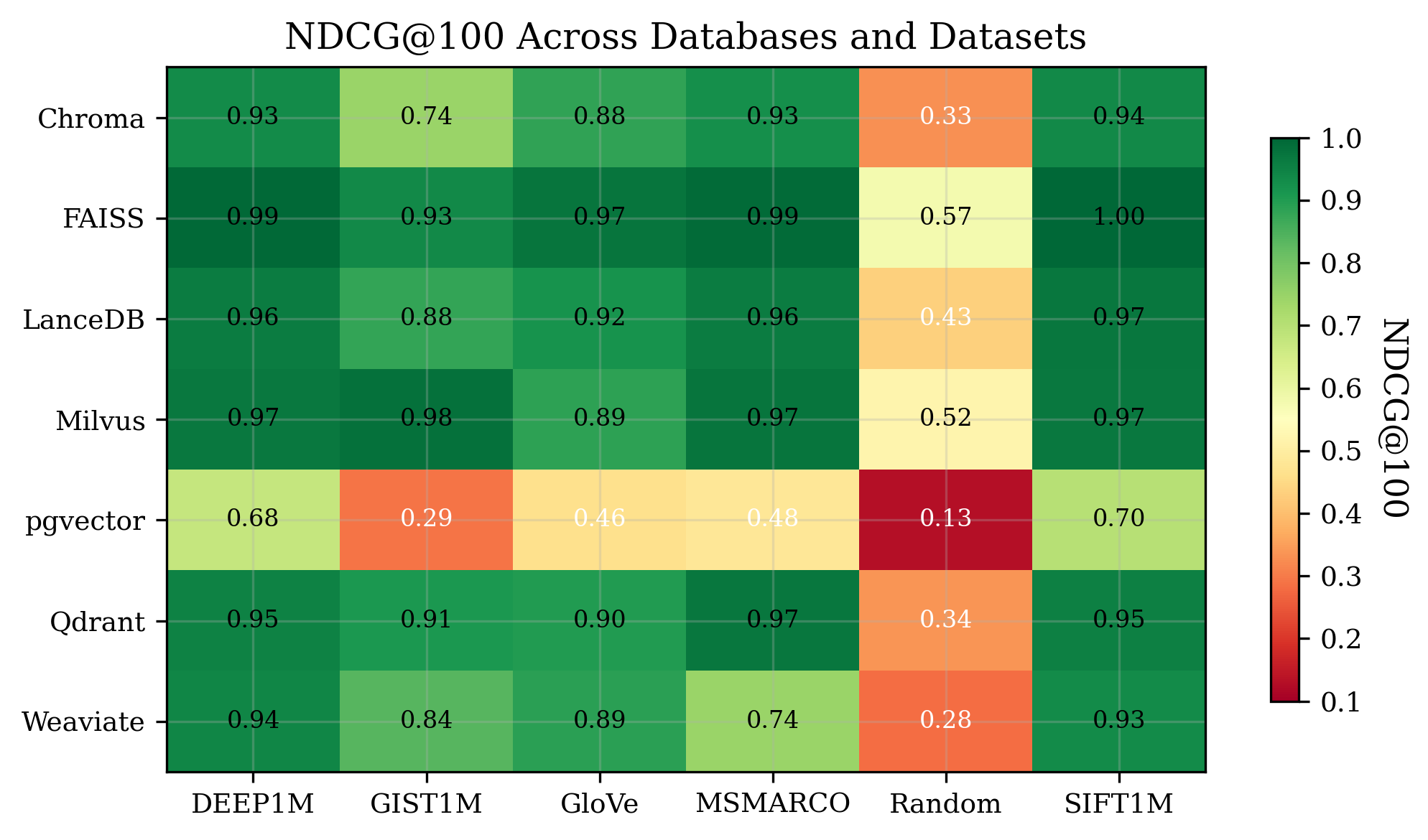}
\caption{NDCG@100 across all database--dataset pairs. Darker green indicates higher ranking quality. FAISS leads overall; pgvector is weakest on Random and GIST1M.}\label{fig:ndcg}
\end{figure}

\begin{figure}[t]
\centering
\includegraphics[width=0.98\linewidth]{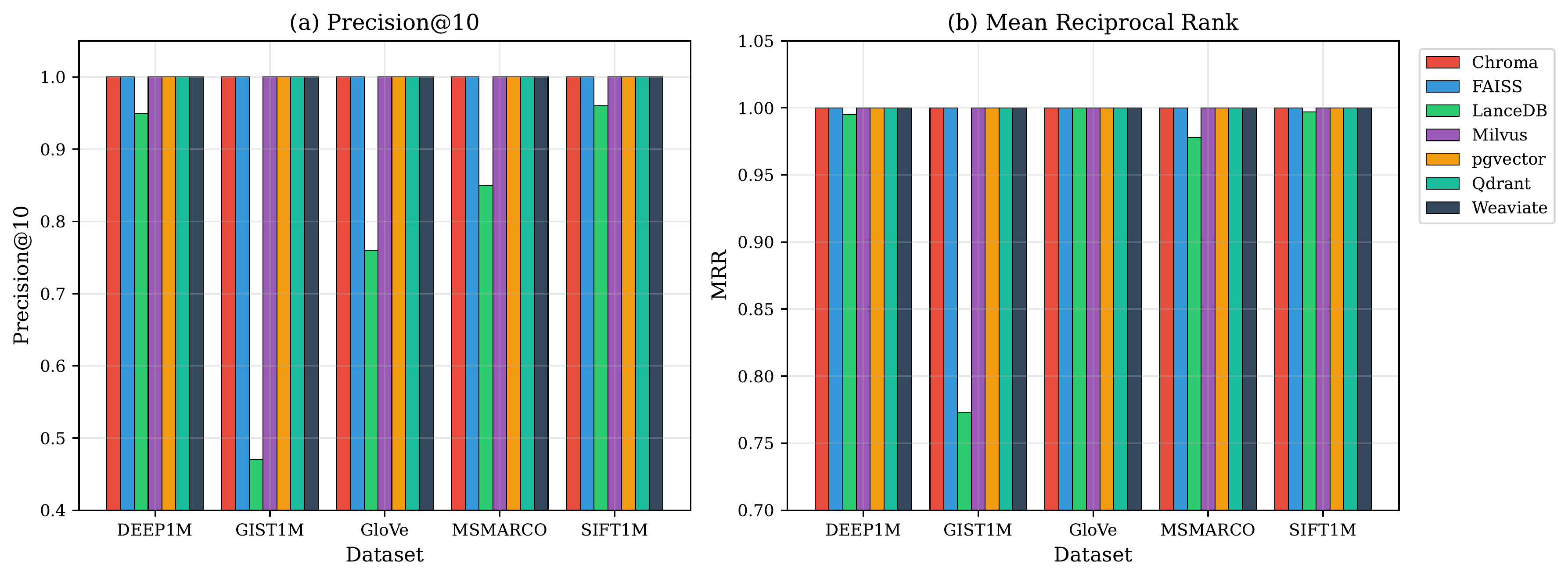}
\caption{(a) Precision@10 and (b) MRR by dataset. Most systems score near 1.0 on both; LanceDB trails on GIST1M.}\label{fig:precmrr}
\end{figure}

Beyond recall, ranking quality determines whether the retrieved neighbors are ordered correctly by true distance. Fig.~\ref{fig:ndcg} reports NDCG@100 as a heatmap across all database--dataset pairs. FAISS attains the highest NDCG on most datasets (0.99 on DEEP1M, 1.00 on SIFT1M) despite not leading on raw recall, indicating that the neighbors it does retrieve are ranked with high geometric precision, a benefit of its optimized, numerically precise distance computation. The heatmap also reveals a consistent per-column pattern: every system's weakest column is Random (values from 0.13 to 0.57), and every system's strongest columns are DEEP1M, MSMARCO, and SIFT1M (mostly 0.9 or above), showing that ranking quality, like recall, is driven primarily by dataset structure rather than by database architecture alone. pgvector shows the weakest ranking quality of any cell in the heatmap on Random and GIST1M (0.13 and 0.29 respectively), consistent with its lower recall on those datasets, while Milvus is the only system to score above 0.97 on GIST1M, mirroring its recall advantage there. Fig.~\ref{fig:precmrr} shows Precision@10 and MRR; most systems achieve near-perfect scores on both (clustering at or above 0.9), with LanceDB the notable exception on the high-dimensional GIST1M dataset, where both metrics visibly dip below the rest of the cohort, reinforcing that its quantization strategy specifically struggles with high-dimensional, sparse feature spaces rather than degrading uniformly across all data types.

\subsection{Query Throughput and Latency}
\begin{figure}[t]
\centering
\includegraphics[width=0.98\linewidth]{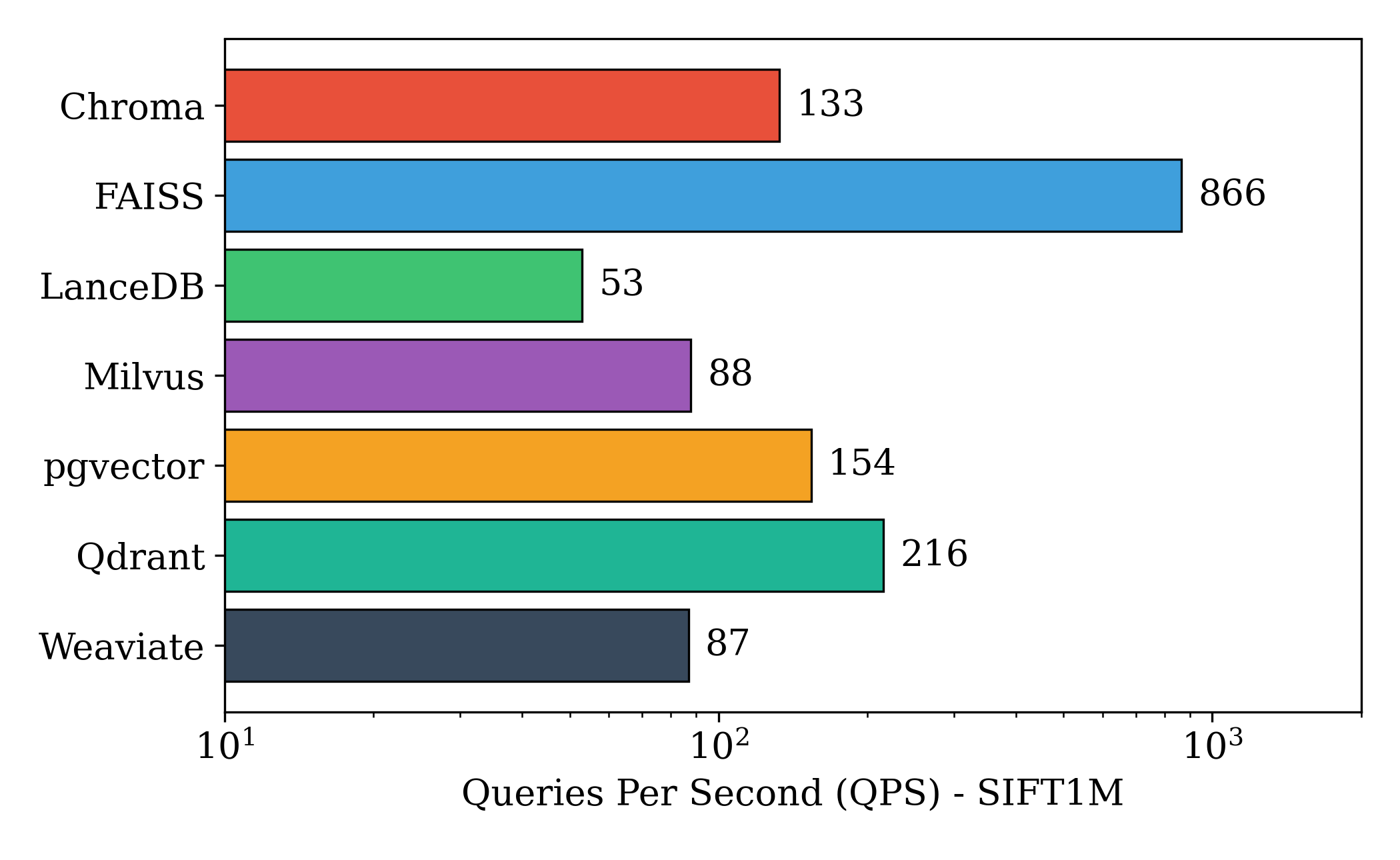}
\caption{Query throughput (QPS, log scale) on SIFT1M. FAISS leads by 4--16$\times$ over full database systems; Qdrant leads among databases.}\label{fig:qps}
\end{figure}

Fig.~\ref{fig:qps} shows single-threaded throughput on SIFT1M. FAISS reaches 866 QPS, 4--16$\times$ higher than the full database systems, because it bypasses network, serialization, and transactional overhead as an in-process library. Among databases, Qdrant leads (216~QPS), followed by pgvector (154), Chroma (133), Milvus and Weaviate (${\sim}$87--88), and LanceDB (53), whose disk-based architecture is the slowest; notably, the two systems with the highest recall in Fig.~\ref{fig:recall} (Weaviate and Milvus) occupy the bottom half of the throughput ranking, an early signal of the recall--throughput tension explored quantitatively in Section~\ref{sec:results}D.

\begin{figure}[t]
\centering
\includegraphics[width=0.98\linewidth]{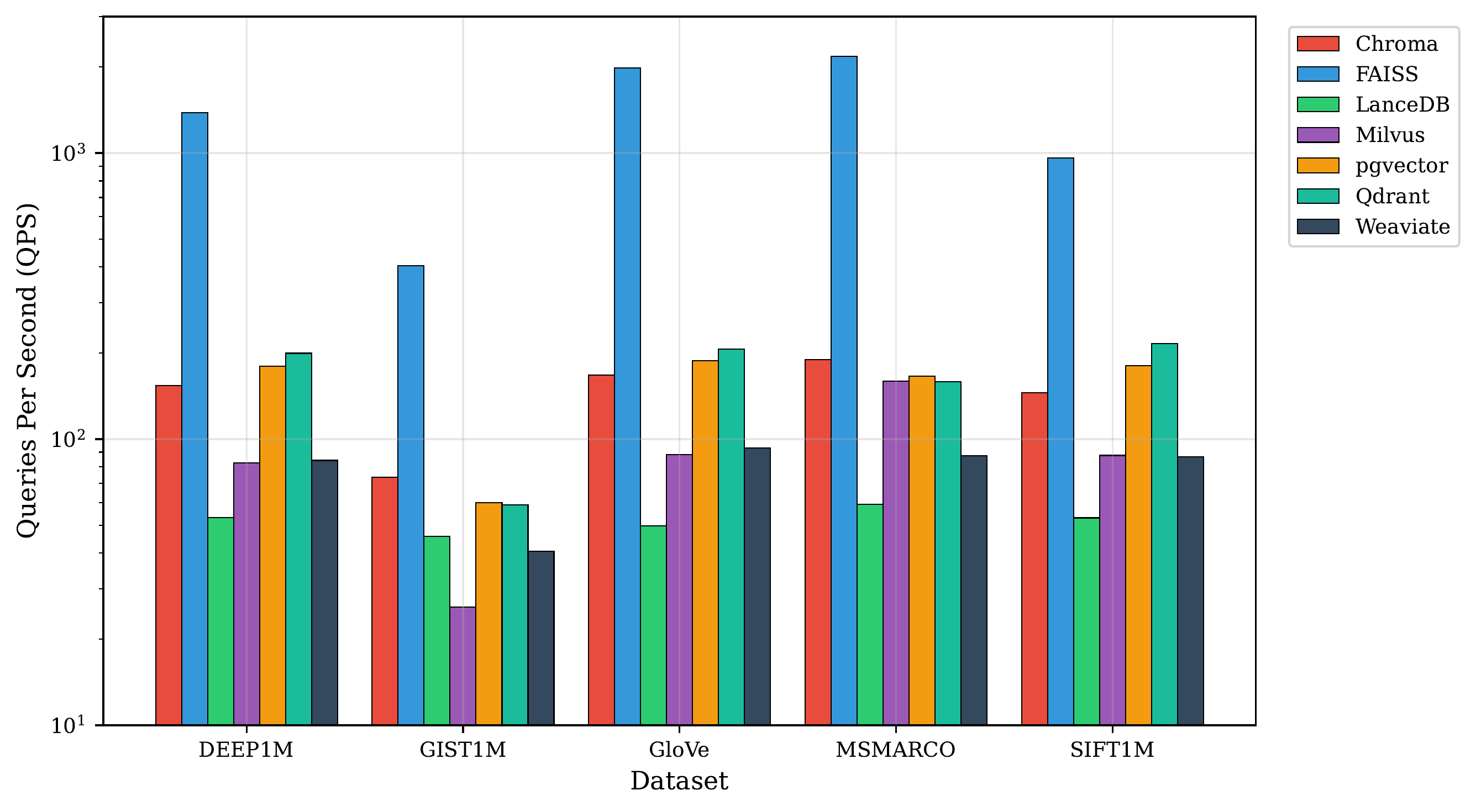}
\caption{Query throughput across all six datasets (log scale). Throughput drops 2--5$\times$ on the 960-dimensional GIST1M relative to lower-dimensional sets.}\label{fig:qpsscale}
\end{figure}

Fig.~\ref{fig:qpsscale} shows throughput across all six datasets. Performance generally degrades with increasing dimensionality: GIST1M (960D) shows 2--5$\times$ lower throughput than the lower-dimensional datasets for every system, confirming that per-query distance computation cost, rather than index type alone, dominates at high dimensionality. FAISS retains the largest absolute margin over the other systems on every dataset, but that margin compresses on GIST1M, where its lead over Qdrant narrows most visibly, suggesting that FAISS's low per-query overhead matters relatively less once raw distance computation over 960 dimensions becomes the bottleneck.

\begin{figure}[t]
\centering
\includegraphics[width=0.85\linewidth]{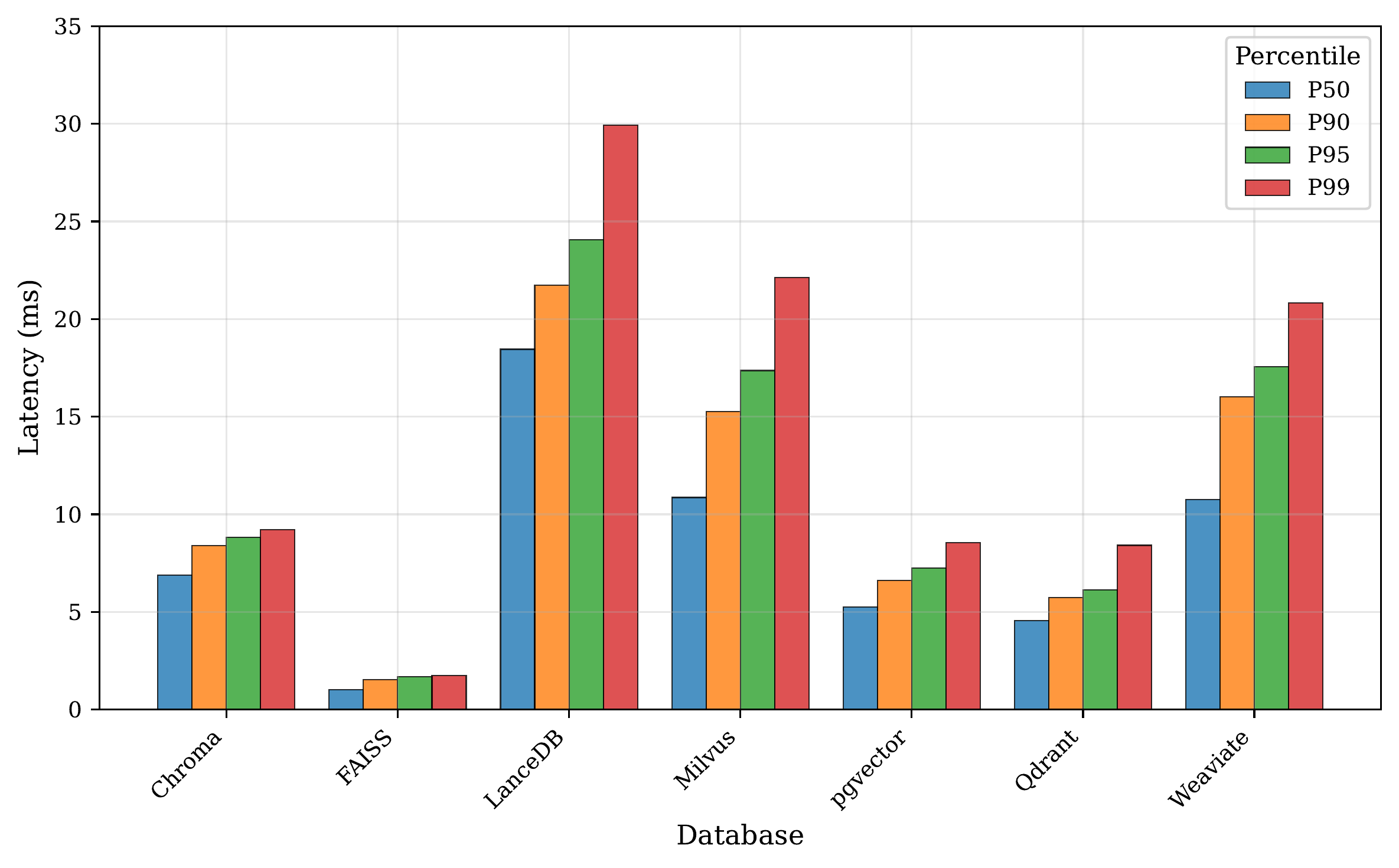}
\caption{Latency percentiles (P50/P90/P95/P99) by database on SIFT1M. FAISS is fastest and most consistent; Qdrant leads among full databases.}\label{fig:latency}
\end{figure}

Fig.~\ref{fig:latency} presents P50/P90/P95/P99 latency on SIFT1M. FAISS sustains sub-2~ms latency at all percentiles, and its bars are visibly the flattest in the chart, the clearest indicator of tail-latency consistency among the seven systems. Among databases, Qdrant is both fastest (4.55~ms P50) and most consistent (P99/P50 = 1.85); pgvector shows the tightest tail ratio (1.63) despite higher absolute latency, meaning its worst-case queries are proportionally closer to its typical query than any other database, a desirable property for SLA-bound deployments even though its median is not the lowest. Milvus (2.04) and Weaviate (1.94) show the largest tail-to-median spread, consistent with their distributed, multi-stage query paths introducing occasional latency spikes not present in simpler single-process architectures.

\subsection{Index Build Time and Resource Usage}
\begin{figure}[t]
\centering
\includegraphics[width=0.98\linewidth]{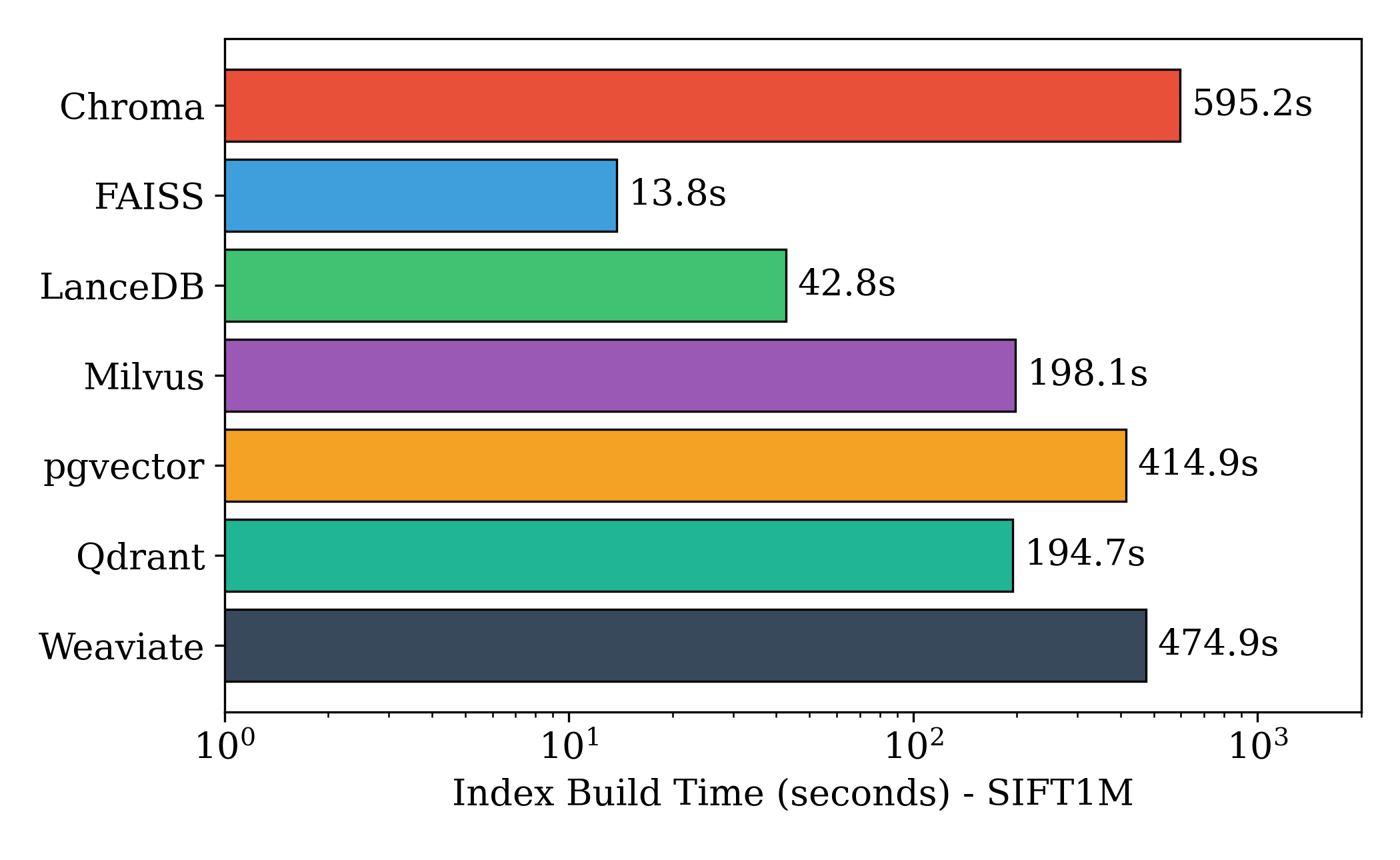}
\caption{Index build time (s, log scale) on SIFT1M. FAISS is fastest; Chroma is slowest, ahead of Weaviate.}\label{fig:build}
\end{figure}

Fig.~\ref{fig:build} shows index construction time on SIFT1M, spanning roughly two orders of magnitude, from 13.8~s to 595.2~s. FAISS builds in 13.8~s via an optimized, SIMD-accelerated C++ implementation. LanceDB follows at 42.8~s, since IVF-PQ construction is inherently cheaper than HNSW graph construction, a $3.1\times$ gap over FAISS that is far smaller than the throughput or recall gaps between the two systems, showing that build speed and query-time behavior are governed by largely independent architectural choices. Qdrant (194.7~s) and Milvus (198.1~s) reflect standard HNSW build cost and are nearly identical, differing by under 2\%, despite belonging to different architectural families (single-node Rust engine vs.\ distributed storage/compute separation); pgvector (414.9~s) adds PostgreSQL transactional overhead, roughly $2.1\times$ Qdrant's build time. Weaviate (474.9~s) and, notably, Chroma (595.2~s) invest the most build time, consistent with higher-quality HNSW parameters; Chroma is the slowest system to index on SIFT1M in our evaluation, exceeding even Weaviate by 25\%, which is a departure from Chroma's comparatively modest recall and mid-tier throughput and suggests its default index configuration is not particularly build-time-efficient relative to the quality it delivers.

\begin{figure}[t]
\centering
\includegraphics[width=0.98\linewidth]{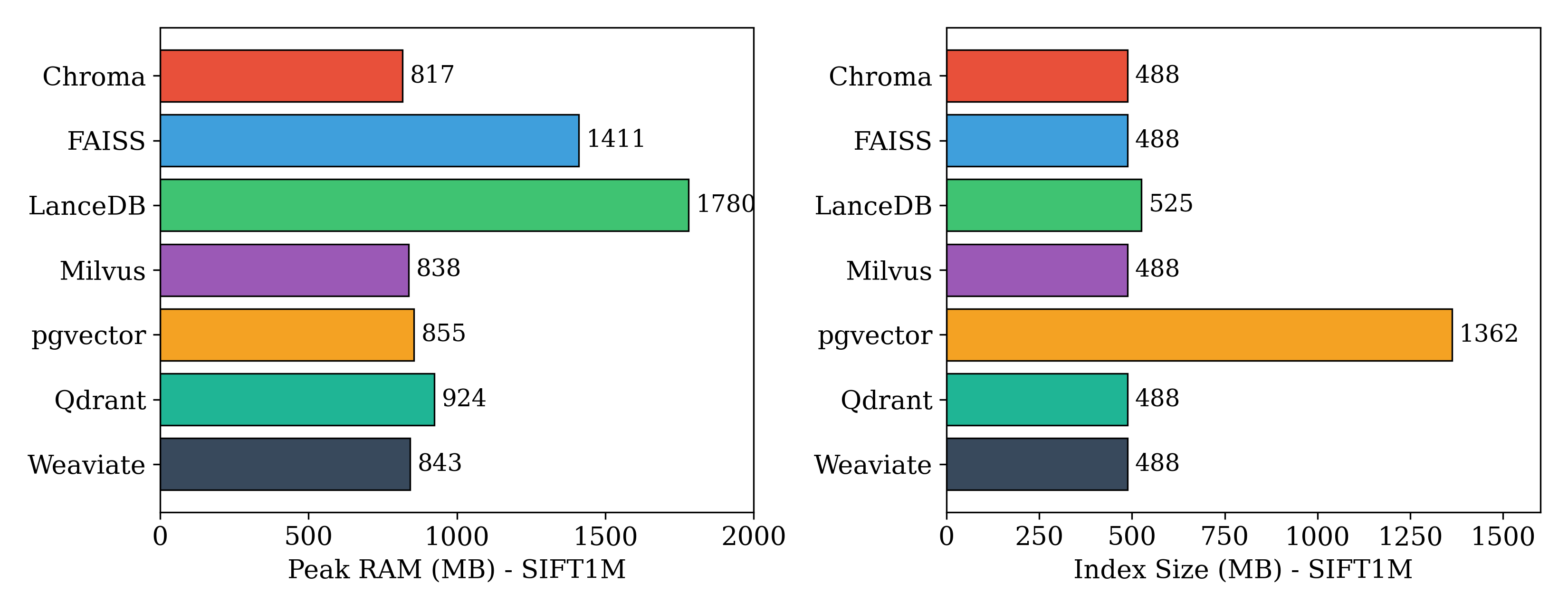}
\caption{Peak RAM (left) and index size (right) on SIFT1M. LanceDB and FAISS lead in RAM; pgvector is the storage outlier.}\label{fig:resources}
\end{figure}

Fig.~\ref{fig:resources} presents peak RAM and on-disk index size. Five systems cluster tightly between 817--924~MB peak RAM (Chroma, Milvus, Weaviate, pgvector, Qdrant), a span of just over 100~MB despite their architectural diversity; LanceDB (1{,}780~MB) and FAISS (1{,}411~MB) are outliers, the former from memory-mapped file access and the latter from an uncompressed in-memory index, together showing that the two ends of the design spectrum, purely disk-oriented and purely in-memory, both carry a RAM penalty relative to the middle-ground database systems. On disk, pgvector requires 1{,}362~MB, roughly 2.8$\times$ the 488--525~MB used by the other systems, due to PostgreSQL's uncompressed row-based storage; every other system converges to nearly the same on-disk footprint (488~MB, with LanceDB slightly higher at 525~MB), indicating that, unlike RAM, on-disk index size is largely architecture-independent once a system moves outside the relational-database paradigm.

\subsection{Trade-off Analysis}
\begin{figure}[t]
\centering
\includegraphics[width=0.98\linewidth]{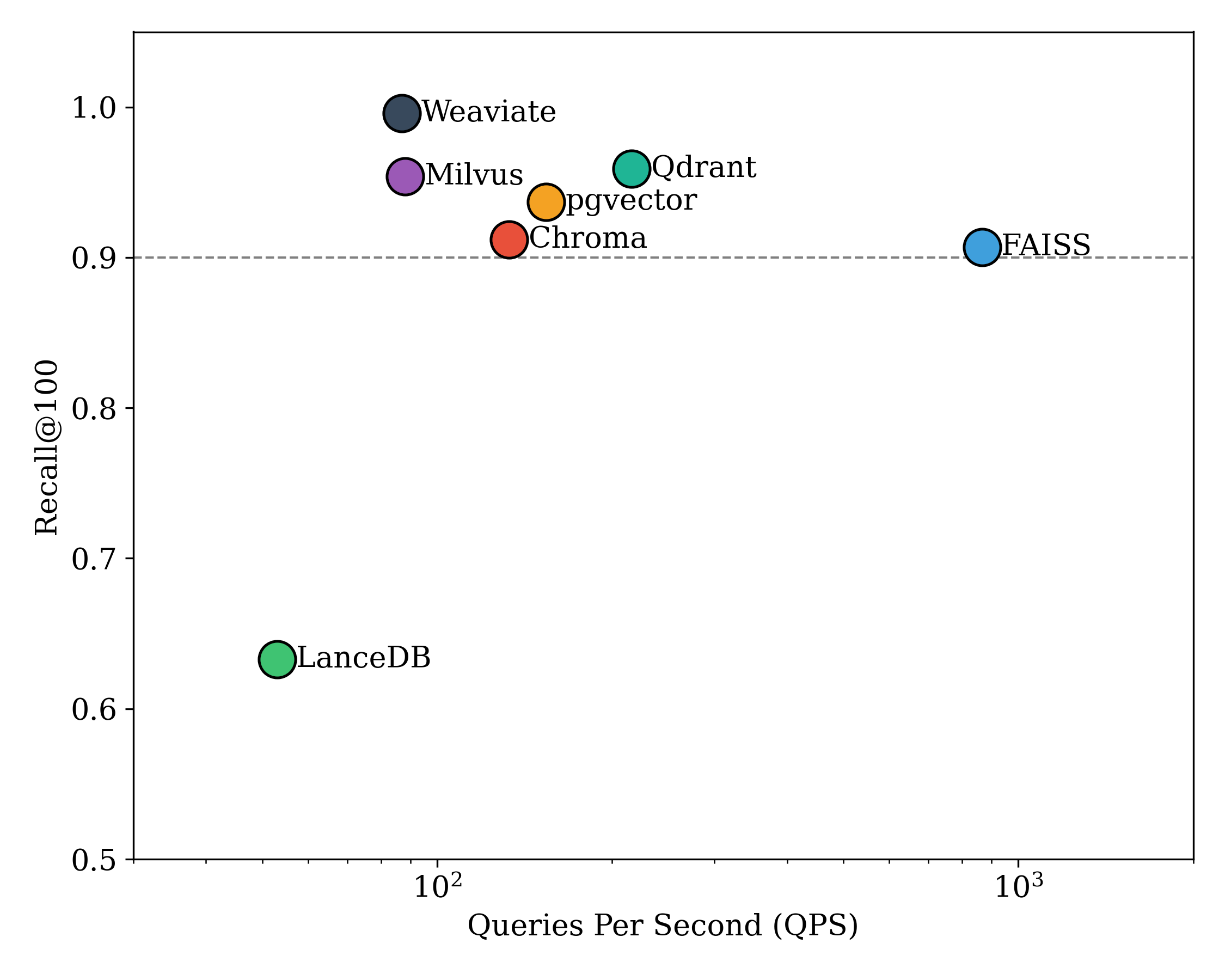}
\caption{Recall@100 vs. QPS on SIFT1M (log scale). Systems separate into high-throughput, balanced, quality-focused, and efficiency-focused regimes.}\label{fig:tradeoff}
\end{figure}

Fig.~\ref{fig:tradeoff} visualizes the recall--throughput trade-off on SIFT1M. Systems form four regimes: a high-throughput regime (FAISS: 866~QPS, 0.907 recall); a balanced regime (Qdrant: 216~QPS/0.959 recall; pgvector: 154/0.937; Chroma: 133/0.912); a quality-focused regime (Weaviate: 87/0.996; Milvus: 88/0.954); and an efficiency-focused regime (LanceDB: 53/0.633). Visually, six of the seven systems sit above the 0.90 recall line despite spanning more than an order of magnitude in QPS, and only FAISS trades a small amount of recall (0.907, just below Weaviate's 0.996) for its outsized throughput advantage, indicating that, on SIFT1M specifically, high throughput does not require a correspondingly large recall sacrifice, LanceDB being the sole exception. The equivalent build-time-versus-recall relationship shows FAISS and LanceDB achieving fast construction at differing recall costs, while Weaviate and Chroma invest the most build time, though Chroma's added cost over Weaviate does not yield higher recall, reinforcing the build-time finding in Section~\ref{sec:results}C that Chroma's indexing cost is not well spent relative to the quality it returns.

\begin{figure}[t]
\centering
\includegraphics[width=0.98\linewidth]{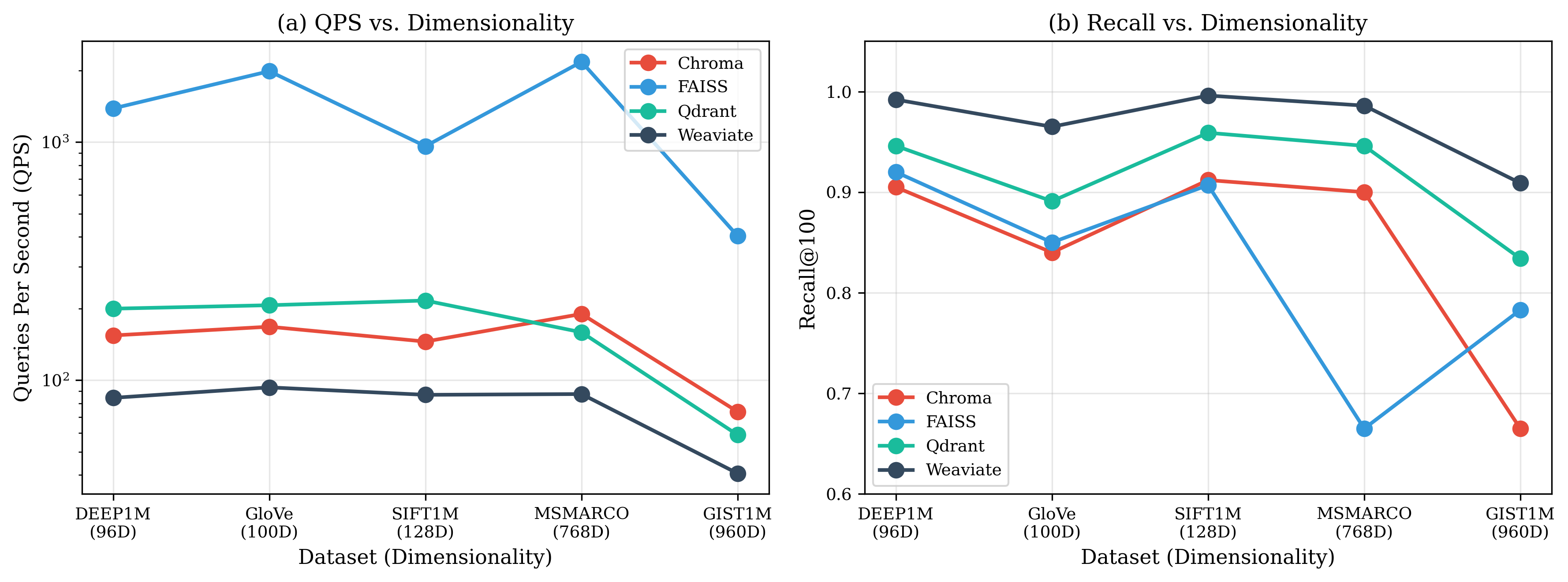}
\caption{Impact of embedding dimensionality on (a) throughput and (b) recall. All systems degrade on the 960-dimensional GIST1M; Milvus degrades least.}\label{fig:dimension}
\end{figure}

Fig.~\ref{fig:dimension} isolates the effect of dimensionality on throughput and recall for the four systems with the widest dimensionality range in our dataset suite. Throughput falls 2--5$\times$ and recall falls 5--15\% on the 960-dimensional GIST1M relative to the lower-dimensional datasets, with Milvus showing the most robust high-dimensional behavior among the systems compared, as its recall curve is visibly flatter across the dimensionality axis than the other three systems shown, which each show a sharper downward inflection at GIST1M.

\begin{figure}[t]
\centering
\includegraphics[width=0.85\linewidth]{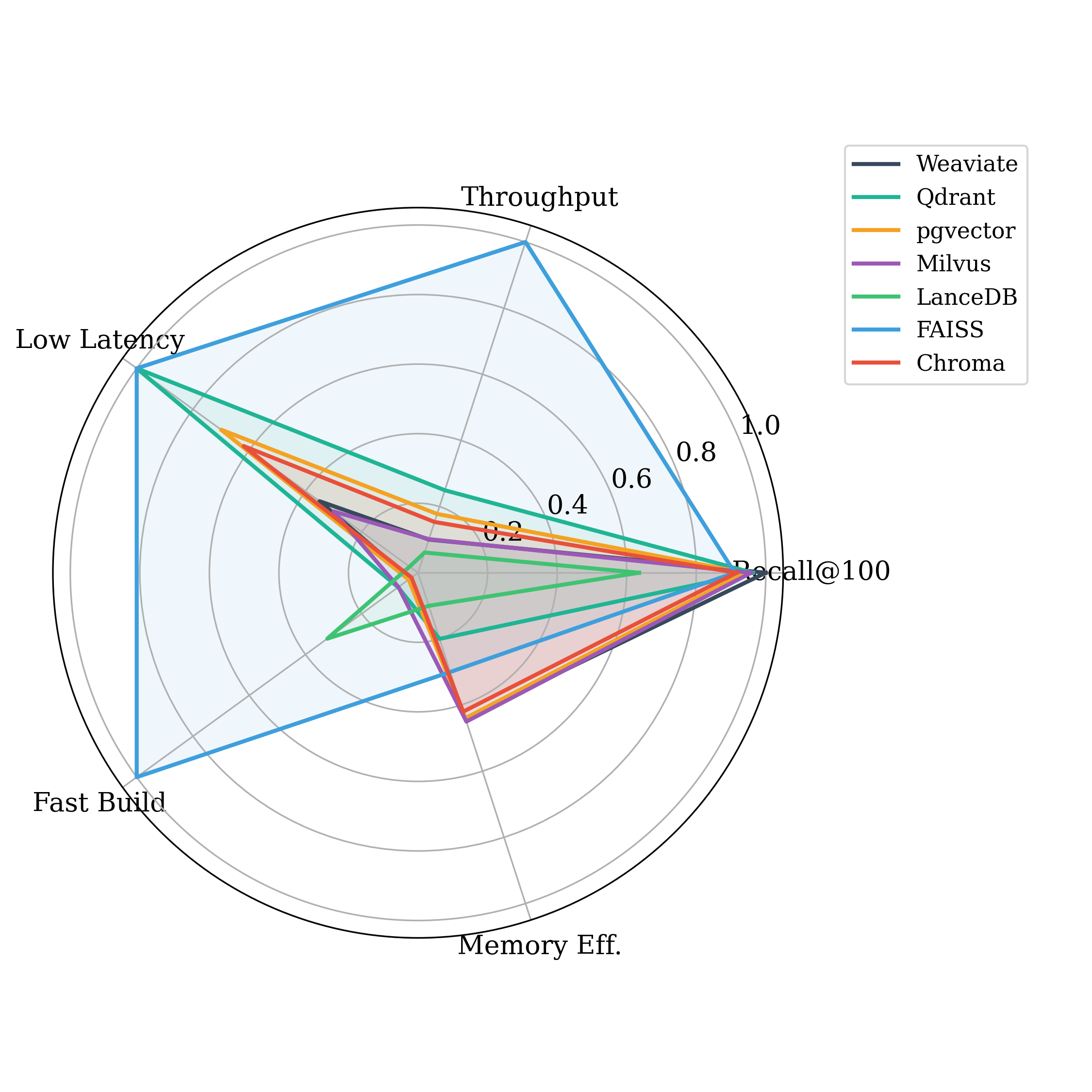}
\caption{Normalized radar comparison across Recall@100, Throughput, Low Latency, Fast Build, and Memory Efficiency (all axes scaled so 1.0 is best-in-class). FAISS dominates three of five axes; no system dominates all five.}\label{fig:radar}
\end{figure}

Fig.~\ref{fig:radar} synthesizes five normalized metrics, Recall@100, Throughput, Low Latency, Fast Build, and Memory Efficiency, into a single view. FAISS forms the most extreme silhouette, reaching the outer edge on Throughput, Low Latency, and Fast Build simultaneously while remaining competitive on Recall@100, visually confirming it as the only system with no clearly weak axis apart from Memory Efficiency. Weaviate, Milvus, and Qdrant form a visually similar, more balanced silhouette clustered toward the Recall@100 vertex, while LanceDB's polygon is the smallest overall, reflecting its comparatively low scores on Recall@100, Throughput, and Low Latency simultaneously; no single non-FAISS system's polygon strictly encloses another's, reinforcing that the choice among the six full database systems depends on which axis a given deployment prioritizes.

\subsection{Limitations}
Results reflect default, untuned configurations, which may understate the ceiling of highly configurable systems such as FAISS and Milvus. The single-node setup does not capture distributed scaling for Milvus or Qdrant, and we evaluate pure vector search without attribute filtering, which may shift relative performance for systems with native filtering support. Hardware specificity (6-core, no GPU) and rapid system evolution further bound the generalizability of absolute figures, though relative architectural trends are expected to persist.

\section{Conclusion}\label{sec:conclusion}

We presented a comprehensive empirical evaluation of seven vector database systems across six datasets and 15 metrics spanning retrieval quality, performance, and resource utilization. No single system dominates every dimension: FAISS leads throughput (866 QPS on SIFT1M, over 2{,}174 on MS~MARCO) and ranking precision but omits database features; Weaviate leads out-of-the-box recall (\textgreater99\%); Qdrant offers the best latency--throughput balance among full databases (4.55~ms P50, 216~QPS); Milvus excels on high-dimensional data (0.971 recall at 960D); pgvector enables SQL-integrated deployment at competitive throughput (154~QPS); LanceDB trades recall for roughly 11$\times$ faster indexing; and Chroma offers a developer-friendly API at the cost of the longest index build time observed (595.2~s). Table~\ref{tab:guidelines} summarizes practical selection guidance derived from these results. These findings, and our open-sourced benchmarking framework, provide practitioners an empirical basis for matching vector database architecture to workload requirements.

\begin{table}[t]
\caption{System Selection Guidelines}\label{tab:guidelines}
\centering
\small
\begin{tabular}{p{1.4cm}p{2.3cm}p{3.6cm}}
\toprule
\textbf{Priority} & \textbf{Recommended} & \textbf{Rationale} \\
\midrule
Throughput & FAISS & 4--16$\times$ higher QPS; needs app-level persistence \\
Balanced production & Qdrant & 216 QPS, \textgreater95\% recall, native filtering \\
SQL integration & pgvector & ACID, hybrid SQL/vector queries \\
Max.\ recall & Weaviate & \textgreater99\% recall on SIFT1M \\
High-dim.\ / scale & Milvus & Best GIST1M (960D) recall; distributed \\
Fast iteration & LanceDB & 11$\times$ faster build than Weaviate \\
\bottomrule
\end{tabular}
\end{table}

\end{document}